\documentclass[aps,12pt]{revtex4}
\usepackage{epsfig}
\usepackage{amsmath,amssymb,color}
\usepackage{epstopdf}
\parskip=\medskipamount

\newcommand{\eq}[1]{(\ref{#1})}
\newcommand{\fig}[1]{Fig.\ref{#1}}

\newcommand{\be}{\begin{equation}}
\newcommand{\ee}{\end{equation}}

\newcommand\disp{\displaystyle}

\newcommand{\la}{\left<}
\newcommand{\ra}{\right>}

\begin{document}

\title{Finite plateau in spectral gap of polychromatic constrained random networks}

\author{V. Avetisov$^{1,6}$, A. Gorsky$^{2,3}$, S. Nechaev$^{4,5}$, O. Valba$^{1,6}$}

\address{$^{1}$N.N. Semenov Institute of Chemical Physics of the Russian Academy
of Sciences, 119991, Moscow, Russia \\
$^2$Institute of Information Transmission Problems of the Russian Academy of Sciences,
Moscow, Russia, \\
$^3$Moscow Institute of Physics and Technology, Dolgoprudny 141700, Russia \\
%$^5$Universit\'e Paris-Sud/CNRS, LPTMS, UMR8626, B\^at. 100, 91405 Orsay, France, \\
$^4$ Interdisciplinary Scientific Center Poncelet (CNRS UMI 2615), Moscow, Russia \\
$^5$P.N. Lebedev Physical Institute of the Russian Academy of Sciences, 119991, Moscow, Russia \\
$^6$Department of Applied Mathematics, National Research University Higher School of Economics, 101000, Moscow, Russia.}

\begin{abstract}

We consider the canonical ensemble of multilayered constrained Erdos-Renyi networks (CERN) and regular random graphs (RRG), where each layer represents graph vertices painted in a specific color. We study the critical behavior in such networks under changing the fugacity, $\mu$, which controls the number of monochromatic triads of nodes. The behavior of considered systems is investigated via the spectral  properties of the adjacency and Laplacian matrices of corresponding networks.
%The cases of fugacities for trimers+triangles and for  triangles only are considered separately.
For some wide region of $\mu$ we find the formation of a finite plateau in the number of the intercolor links, which exactly matches the finite plateau for the algebraic connectivity of the network (the value of the first non-vanishing eigenvalue of the Laplacian matrix, $\lambda_2$). We claim that at the plateau the restoring of the spontaneously broken $Z_2$ symmetry by the mechanism of modes collectivization in clusters of different colors occurs. The phenomena of a finite plateau formation holds for the polychromatic (multilayer) networks with $M>2$ colors.

\end{abstract}

\maketitle

\section{Introduction}

Critical phenomena in topological random networks and random regular graphs are under scrutiny in past decade. One of the most intriguing phenomena is the clusterization happen in dynamical (evolutionary) networks when some control parameters are changing. In unconstrained Erdos-Renyi random graphs the clusterization is known as the Strauss transition and occurs when the evolving network tends to increase the number of closed triads of bonds, $n_{\triangle}$. It is convenient to deal with the canonical ensemble, in which the value $n_{\triangle}$ is coupled to the fugacity, $\mu$, playing the role of the "energy" assigned to each closed triad of bonds. At some critical value, $\mu_{cr}$, the network falls into the "Strauss phase" -- a single full sub-graph with a maximally possible number of triangles. This phenomena has been studied and very satisfactory described in the frameworks of the random matrix model \cite{burda}, and using the mean-field arguments \cite{newman1}.

The detailed analysis of the phase transition in constrained Erdos-Renyi networks (CERN) and in random regular graphs (RRG) carried in \cite{decay}, amounts to the picture which is essentially different from the Strauss clusterization. It has been found that the condition of maximization of $n_{\triangle}$, forces the random network with the conserved vertex degree to form a multi-clique ground state (contrary to a single-clique ground state in the Strauss model). The number of cliques asymptotically equals $[1/p]$ for CERNs, where $Np$ is the average vertex degree at the network preparation conditions, while for the RRGs the maximal number of emerging cliques is $[N/k]$, where $k$ is the vertex degree. In terms of the spectrum of the adjacency matrix, the decay looks as the tunneling of $[1/p]$ isolated eigenvalues corresponding to the discrete part of spectrum, from the continuum part of the spectrum to the second nonperturbative zone. This second zone corresponds in the matrix model language to the filling of the maximum of the effective potential for the eigenvalues, while the main continuum zone corresponds to the filling of the absolute minimum of the potential. The $U(N)$ symmetry of the initial state gets broken into the product of smaller symmetry groups corresponding to the block structure of the ground state adjacency matrix.

Considering the Laplacian matrix, $L$, instead the adjacency one, $A$, we can check that the non-perturbative zone of $L$ is formed by "soft" modes, which tunnel from the "hard" part of the spectrum. Recall that the Laplacian matrix of the graph is related to the adjacency matrix as follows:
\be
L= d I - A
\ee
where $I$ is  diagonal matrix, and $d$ is the vector of all vertex degrees. For the RRGs, the spectra of $L$ and $A$ are identical, however hard Laplacian modes get mapped onto soft adjacency modes, and vise versa. The Laplacian matrix is positively defined and has the minimal eigenvalue $\lambda_1=0$ corresponding to the homogeneous eigenvector  $\mathbf{v}_1=(1,\dots,1)$. The degree of the degeneration of the Laplacian mode, $\lambda_1$, defines the number of disconnected components of the graph. The behavior of the second eigenvalue of the Laplacian, $\lambda_2$, in RRG is the subject of the several mathematical studies \cite{second}. It has an important meaning for any network known as "the algebraic connectivity". In particular, $\lambda_2>0$ when the graph is connected. The value of $\lambda_2$ plays an important role in the relaxation and transport properties of the network, since it defines the inverse diffusion time, and is crucial for determining synchronization of multiplex networks. The corresponding eigenvector (Fiedler vector) establishes the bijection between the layers of the network.

Two very important properties of the spectral density of CERN and RRG adjacency matrices above the critical point have to be mentioned. First, it was found in \cite{decay} that the spectrum of each clique and the spectrum of the whole network are very different. The spectrum of the clique exhibits the discrete structure typical for the sparse-like graphs \footnote{It is known \cite{tao} that the spectral statistics of sparse-like and almost complete random graphs coincide up to the shift along the eigenvalue axis}, while the spectrum of the whole network has two-zonal structure with the continuous triangle-like form of the first (central) zone. We have interpreted the presence of the second zone as the collectivization (or synchronization) between the modes in different clusters. The second property concerns the non-ergodic behavior of modes in the central zone of the spectrum. It was found in \cite{anderson} that there is a memory on the initial (preparation) conditions in the continuum part of spectrum, which is a signature of the non-ergodic behavior and existence of some hidden conservation laws. In the nonperturbative (second) zone all modes in CERN are localized \cite{anderson}, while the ones in the continuous (central) zone remain delocalized. Qualitatively the same behavior has been seen for RRG as well.

A bunch of new critical phenomena has been found in \cite{color} for "polychromatic" CERN and RRG, for which it is implied that vertices are painted (once and for all) in two colors, "black" and "white", at the preparation conditions. Instead of maximizing the number of closed triads of bonds, we maximize the number of unicolor trimers (irrespective for a moment on the trimer topology). In the canonical description, we associate the fugacity (the energy) $\mu$ to every unicolor (both, black and white) trimer \footnote{We use for the fugacity of unicolor trimers the same notation, $\mu$, as for the fugacity of closed triples of bonds in monochromatic networks}. It turns out that initially $Z_2$-symmetric network is absolutely unstable and undergoes the spontaneous breakdown at any value $\mu>0$ of unicolor trimer fugacity. Therefore we can consider the bichromatic network with broken $Z_2$ symmetry as an example of a two-layer network -- see \cite{arenasrev, boca} for the overviews of multilayer ("multiplex") networks. We have found in \cite{color} that at some critical value, $\mu_{cr}$, the number of black-white links undergoes a phase transition and develops a wide, though \emph{finite} plateau. Some possible interpretations of the plateau formation have been suggested in \cite{color}, however more involved analysis is required to clarify this issue.

Here we discuss the spectral properties of the adjacency and Laplacian matrices of polychromatic networks. Our aim is to determine the spectral density in the ground state of the perturbed CERN and RRG focusing at the statistics of highest eigenvalues of the adjacency matrix and, correspondingly, lowest eigenvalues of the Laplacian matrix. We shall separately investigate the models with fugacities associated with closed triangles of bonds (for colorless networks) and with unicolor trimers of nodes (for polychromatic networks). The case of fugacities for unicolor triangles in polychromatic network can be considered as the generalization of the colorless triangles pattern. The behavior of polychromatic network in this case is similar to the one of monochromatic one -- the network gets defragmented into monochromatic cliques above some critical fugacity.

Our findings for the polychromatic network are as follows:
\begin{itemize}
\item The bichromatic (polychromatic) network is absolutely unstable with respect to the fugacity of unicolor triads of nodes, $\mu$, and immediately splits for any $\mu>0$ into the collection of monochrome weakly connected sub-graphs-cliques, or "layers" (one layer -- one color). From such a viewpoint, $\mu$ is the energy of the simplest \emph{in-layer} triple vertex interaction.
\item In a wide interval, $\mu\in[\mu_{in}, \mu_{out}]$ the number of inter-cluster connections (i.e. cross-color bonds) stays unchanged. For the bichromatic network this behavior is identical to the plateau formation in the second eigenvalue, $\lambda_2(\mu)$, of the corresponding Laplacian matrix: $\lambda_2(\mu)=\mathrm{const}$ for $\mu\in[\mu_{in}, \mu_{out}]$.
\item The plateau is finite, being the synchronization region of monochromatic layers (cliques). Between the values $\mu_{in}$, at which the plateau begins, and $\mu_{out}$, at which it ends, the layers are synchronized and we can say on the effective  restoration  of $Z_2$-symmetry for bichromatic networks (and $Z_M$-symmetry for $M$-color networks).
\end{itemize}

The most interesting question in statistics of multilayer networks, concerns the influence of the interlayer interaction on the properties of the whole network. The first striking example of the dramatic effects enforced by interlayer interactions, has been found in the percolation context in \cite{buldyrev}. Later it was recognized that local interactions in the network affect dramatically diffusion properties of the whole system \cite{arenas, vanmighem, radicchi}. In \cite{arenas} the authors found the formation of the semi-infinite plateau for the second eigenvalue $\lambda_2$ of the Laplacian matrix in a two-layer network with specific inter-layer interaction. If was argued that depending on the strength of interaction, two network layers can behave coherently or individually. The corresponding property can be read off from the eigenvectors of lowest Laplacian eigenvalues.

Key differences between the setup discussed in \cite{arenas,vanmighem,radicchi} and results of our work, are as follows:
\begin{itemize}
\item In our study the two-layer (multi-layer) network emerges dynamically from initially homogeneous polychromatic network, due to simplest color-sensitive trimer interaction, while in \cite{arenas} the layers are handmade.
\item In our work the effect of the plateau formation is due to two facts: the conserved vertex degree of the network, and the in-layer (i.e. the monochromatic) trimer interactions, while in \cite{arenas} the authors control interaction between layers.
\item In our work the plateau is finite with the well specified entrance and exit points, while in \cite{arenas} the plateau for $\lambda_2$ is semi-infinite and apparently cannot be made finite in the frameworks of the considered model.
\end{itemize}

The paper is organized as follows. In Section 2 we briefly describe the models under consideration and analyze the corresponding ground states with introduced fugacities of the unicolor triangles. It is shown that the network decays above the critical fugacity into the cliques similar to \cite{decay}. In Section 3 we consider the networks with fugacities for the unicolor trimers. The evolution of the spectrum yielding the plateau formation and termination is investigated. It is argued that the key phenomena for the plateau formation is the rearrangement of the lowest eigenvalues of the Laplacian. In Section 4 we discuss the localization properties of the spectrum of polychromatic perturbed networks. In Section 5 we compare the phenomena of the plateau formation in our study with previous considerations. The comments on the  related issues and the open questions are summarized in the Discussion.

\section{Models with Chemical potential for triangles}

%\subsection{Generalities}

The conventional colorless Erdos-Renyi (ER) network is prepared by random joining any pair of vertices with the probability $p$. This produces the Poissonian distribution for the vertex degree with the mean value $\la n_{vd} \ra=pN$. For the ensemble of ER networks  one can introduce the spectral density of the adjacency matrix:
\be
\rho(\lambda)= \la \sum_{i=1}^N \delta(\lambda - \lambda_i)\ra
\ee
where $\lambda_i$ is the eigenvalue of the adjacency matrix and, $\la ...\ra$ means the ensemble average. At large $N$ and finite $p$ the spectral density of the ER network adjacency matrix shares the Wigner semicircle distribution
\be
\rho_{\rm W}= \sqrt{c - \lambda^2}
\ee
The isolated largest eigenvalue $\lambda = pN-1$ corresponds to the homogeneous eigenvector
$v=(1,\dots,1)$.

The RRG is the graph with the same number of the links, $k$, at each vertex. Locally the RRG behaves as a Bethe tree, however differs from the infinite Bethe lattice by the presence of cycles of the finite lengths, $\ell$. The spectral density of RRG adjacency matrix shares the Kesten-McKay law \cite{kesten,mckay}
\be
\rho_{\rm KM}(\lambda)= \frac{k\sqrt{4(k-1) - \lambda^2}}{2\pi (k^2 - \lambda^2)}, \qquad
-2\sqrt{k-1}\leq \lambda \leq 2\sqrt{k-1}
\ee
with one isolated eigenvalue located at $\lambda=k$ beyond the main support of the spectral density. The spectral density $\rho_{\rm KM}(\lambda)$ differs from the Wigner semicircle, however tends to it asymptotically at large $k$ upon proper scaling. The finite size correction for the Kesten-McKay law has been found in \cite{parisi}.

The ER networks formed with the probability $p$ and RRG with the vertex degree $k=pN$, have common static properties in many respects
\cite{cycles,spacing} (see \cite{wormald} for review). It particular, it was proved rigorously that the RRG can be "sandwiched" between two very close ER networks and share their statistical characteristics, like cycle distributions and the chromatic numbers \cite{huang}. However, the stochastic dynamics on ER and RRG (as, for instance, the redistribution of links), leads to essentially distinct final states in ER and RRG since the stochastic dynamics is very sensitive to the absence or presence of the vertex degree conservation. In order to make the behavior of two ER and RRG similar at the dynamical level, we impose the degree conservation condition on ER network which leads us to consideration of CERN.

Here we describe the model of dynamically rebuilt polychromatic (bichromatic) network generalizing the setup of the colorless model considered in \cite{decay}. In main steps this construction reproduces the one described in \cite{color}, however some details of the model are more specified, so it seems instructive to provide here the full definition of the system under consideration.

Since each monochromatic closed triple of nodes (the triadic "motif") is weighted with the fugacity $\nu$, the partition function of the system can be written as follows:
\be
Z(\mu) = \sum_{\{\rm states\}} \hspace{-0.25cm} {\vphantom{\sum}}' e^{-\mu n_{\triangle}}
\label{eq:04}
\ee
where prime in \eq{eq:04} means that the summation runs over all possible configurations of nodes, under the condition of fixed degree in each vertex, $i$ ($i=1,..., N$) of the graph.

To reach the true ground state, we run the stochastic evolution of the CERN or RRG (the discrete version of the Langevin equation), from the initial configuration to the ground state. The initial state of the network is prepared by connecting any randomly taken pair of vertices with the probability $p$. Then, one randomly chooses two arbitrary links, say, between vertices $i$ and $j$ ($i$--$j$) and between $k$ and $m$ ($k$--$m$), and reconnect them, getting new links ($i$--$k$) and ($j$--$m$). Such reconnection conserves the vertex degree \cite{maslov}. Now one applies the standard Metropolis algorithm with the following rules: i) if after the reconnection the number of node triples is increased, a move is accepted, ii) if the number of node triples is decreased by $\Delta n_{\triangle}$, or remains unchanged, a move is accepted with the probability $e^{-\mu \Delta n_{\triangle}}$. Then the Metropolis algorithm runs repetitively for large set of randomly chosen pairs of links, until it converges, as it is shown in \cite{algor}, to the true ground state in the equilibrium ensemble of random undirected Erd\H{o}s-Renyi networks with fixed vertex degree.

We maximize the number of unicolor trimers (irrespective for a moment on their topology). In the canonical description, we associate the fugacity $\mu$ to every unicolor (both, black and white) trimer of a large $N$-vertex networks with conserved vertex degree. There are two types of networks with such property: constrained Erdos-Renyi networks (CERN) and random regular graphs (RRG), which demonstrate similar clusterization behavior. At the critical value of the chemical potential, $\mu_{cr}$, the polychromatic network decays into the maximally possible number of the monochrome cliques.

\begin{figure}[ht]
\centerline{\includegraphics[width=15cm]{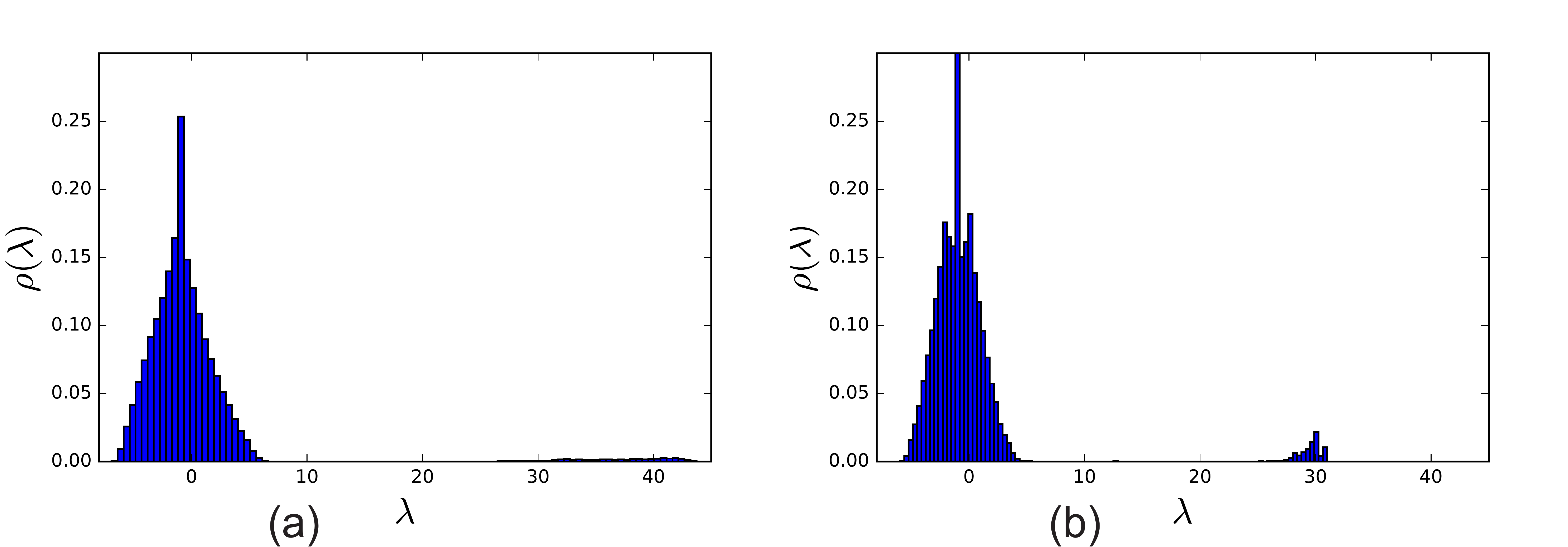}}
\caption{Spectral density in the ground state of the multicolor constrained Erdos-Renyi networks ($p=0.15$) (a) and random regular graphs ($k=38$) (b) above the critical fugacity for unicolor triangles. The numerical results are obtained for the ensembles of 250 random graphs of 256 vertices.}
\label{fig:01}
\end{figure}

The number of the cliques of distinct colors can vary, being dependent on the numbers of vertices of different colors in the initial network. The total spectral density of the ensemble develops two zones, as in the colorless network, see \fig{fig:01}. The non-perturbative zone consists of clusters of various colors. Each cluster is a highly connected clique and its spectral density has a shape typical for sparse matrices, however due to the inter-clique interactions, the fast modes inside the clique get collectivized and the total shape in the central zone acquires the triangle-like shape.

\section{Finite spectral gap plateau  for model with chemical potential for trimers}

\subsection{Numerics}

As one can see, at any value  $\mu>0$ of the chemical potential of monochrome triples, the two-color network undergoes the color separation into two monochrome clusters \cite{color}. So, one can say that after the color separation, the network automatically splits into a two-cluster graph with different weights for inter-layer and in-layer bonds. This splitting does not depend on the initial condition and the two-cluster state is the true ground state of the network at $\mu$.

The important feature of this model deals with the specific plateau formation in the number of black-white links above the critical value $\mu_{cr}$ of the chemical potential of monochrome trimers \cite{color}. The origin of this phenomena remained a bit mysterious in spite of the conjectures made in \cite{color}. It is natural to attack this puzzle via spectral analysis as in the monochromatic case focusing at the spectral gap. We introduce the spectral density of the adjacency matrix for the CERN and RRG ensembles, which does not feel the color structure of the graph, though is sensitive to the cluster formation. We assume that $\mu_b=\mu_w$, hence the Hamiltonian is $Z_2$ symmetric subject to the constraint on the fixed sum of links at each vertex.

\begin{figure}[ht]
\centerline{\includegraphics[width=15cm]{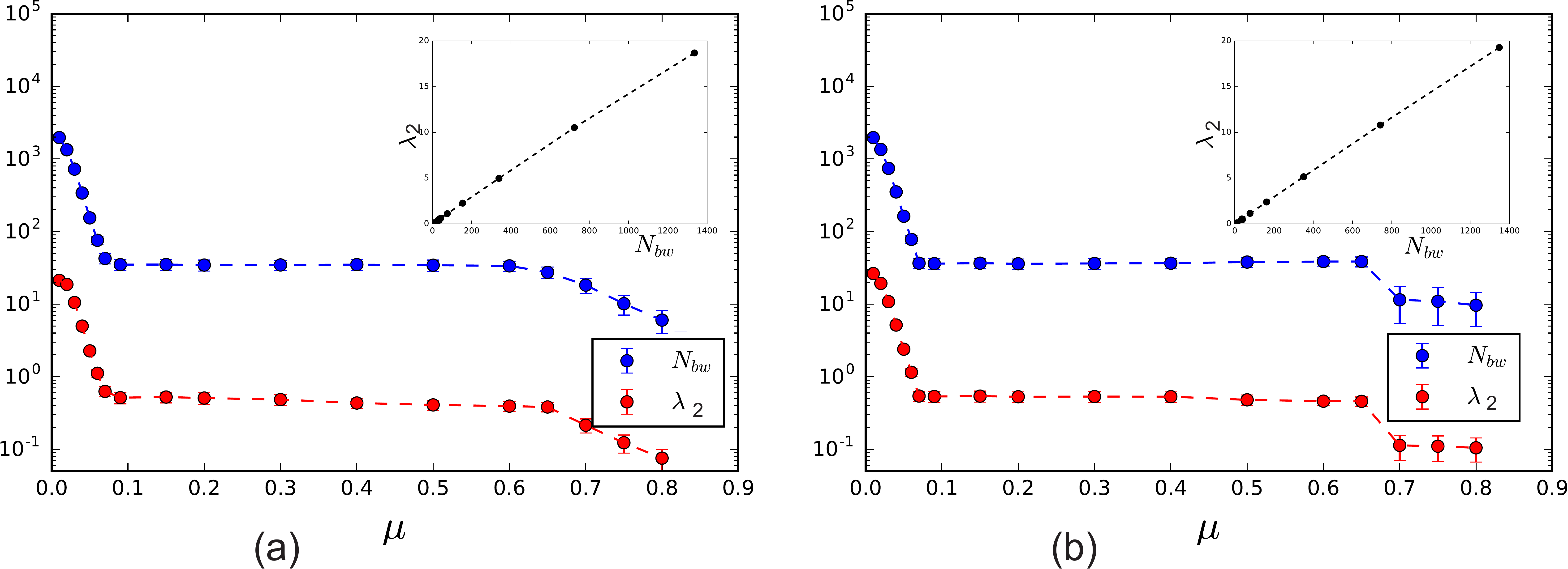}}
\caption{Evolution of the spectral gap in the two-color CERN (a) and RRG (b) versus evolution of number of black-white links. Inserts: the linear dependence of the spectral gap on the  number of black-white links.  The numerical data are obtained by averaging over 500 randomly generated graphs of 256 vertices and with the probability $p=0.15$ for CERN ensemble, and $k=38$ for RRG.}
\label{fig:03}
\end{figure}

At $\mu=0$ the spectral density has the Wigner semicircle and the single isolated eigenvalue. The initial color splitting can be seen in the spectral density as the tunneling of the single isolated eigenvalue outside of the central zone. This happens at very small value of $\mu$. One could say that $Z_2$ symmetry is broken at this stage. When $\mu$ is increasing, isolated eigenvalues $\lambda_1=0, \lambda_2$ move towards each other and merge -- see \fig{fig:03}.

At some critical value, $\mu_{cr}$, we see in \fig{fig:03} two effects happen simultaneously: (i) beginning of the plateau in the spectral gap measured by the algebraic connectivity, $\lambda_2$, of the Laplacian matrix, and (ii) beginning of the plateau in the number of black-white links, $N_{bw}$. The similarity in the behaviors of $\lambda_2$ and of $N_{bw}$ is not occasional: the algebraic connectivity is proportional to the number of links to be cut to get two disconnected components of the graph. Hence, for our network we have
\be
\lambda_2(\mu)= c N_{bw}(\mu)
\ee
where $c$ is come constant. The plateau in the algebraic connectivity is the spectral counterpart of the phenomena described in \cite{color}. The constraint of the degree conservation is crucial for the plateau formation phenomena, and no plateau is formed if one considers the conventional evolution of ER network without vertex degree conservation.

The behavior of $M$-color CERN and RRG with equal chemical potentials for all colors is similar to the one of two-color case. Again, we introduce the fugacity, $\mu$, for monochromatic trimers of vertices. At small value of chemical potential the network is fragmented into $M$ monochromatic clusters which implies the spontaneous breaking of $Z_M$ discrete symmetry. The network acquires the $M$-layer (multiplex) structure. The monochromatic clusters $1,...,M$ are connected by some number of "polychromatic links" which can be considered as the order parameters of the model.  Denote by $N_{ik}$ the number of polychromatic links between the layers $i$ and $k$ . At some value, $\mu_{cr}$, the plateaus in the dependencies $N_{ik}(\mu)$ get formed simultaneously for any pair $(i,k)$. From the viewpoint of the spectrum of the adjacency matrix of CERN and RRG,  $M$ smallest eigenvalues of the Laplacian matrix simultaneously develop plateaus for each pair $\lambda_i, \lambda_k$, where $(i,k) = 1,...,M$.

\begin{figure}[ht]
\centerline{\includegraphics[width=15cm]{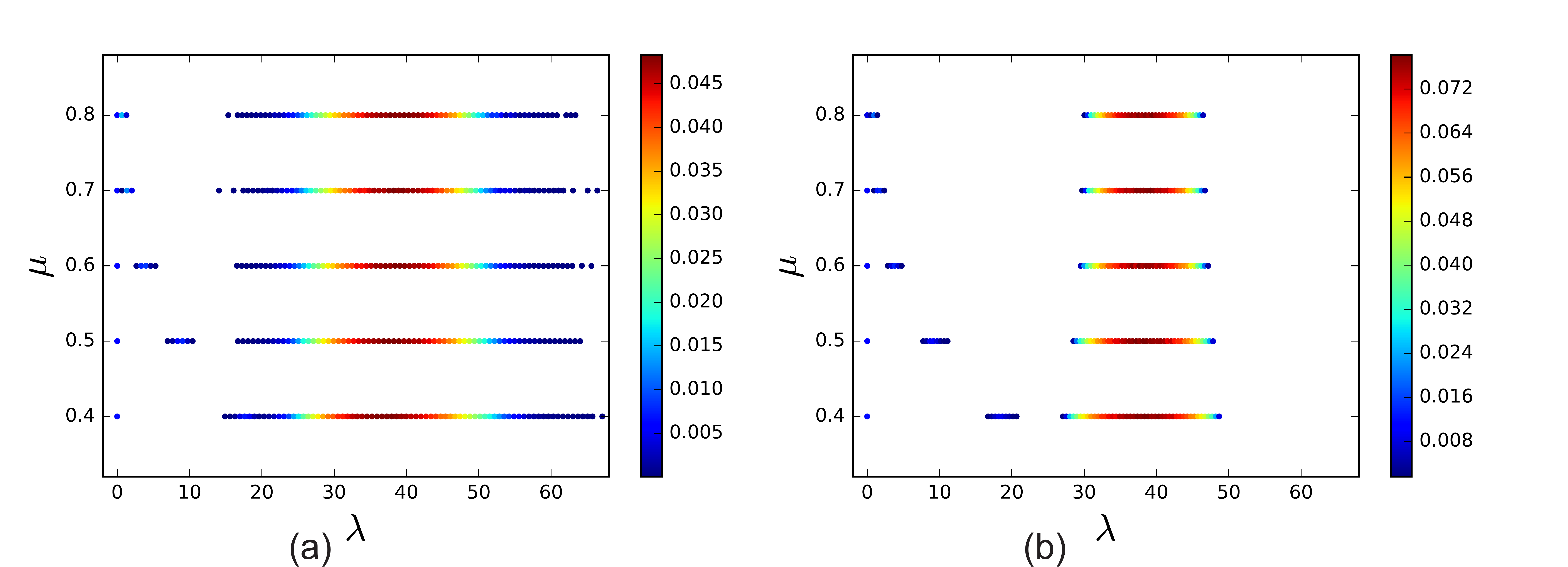}}
\caption{Two-zonal structure for the ensembles of 4-color CERN (a) and RRG (b) upon the color separation. The numerical data are obtained for ensembles of 250 randomly generated graphs of 256 vertices and with the probability $p=0.15$ for CERN, and $k=38$ for RRG.}
\label{fig:04}
\end{figure}

Now we turn to the description of the phase structure of spectral densities in polychromatic ensembles of CERN and RRG. Naively, there are two possibilities: i) isolated eigenvalues corresponding to each color form their "own" non-perturbative zone, or ii) the common second zone involving the eigenvalues of all colors gets formed. The result of the numerical simulations for the case of equal fugacities for all colors is presented in \fig{fig:04}. It demonstrates that the common two-zonal spectral structure plus one isolated eigenvalue $\lambda_1=0$  gets formed. The numerical simulations show that in some interval of fugacities $\mu$, there is plateau for the edge formation of the second non-perturbative zone in the spectral density of ensemble of the Laplacian matrices.

\begin{figure}[ht]
\centerline{\includegraphics[width=15cm]{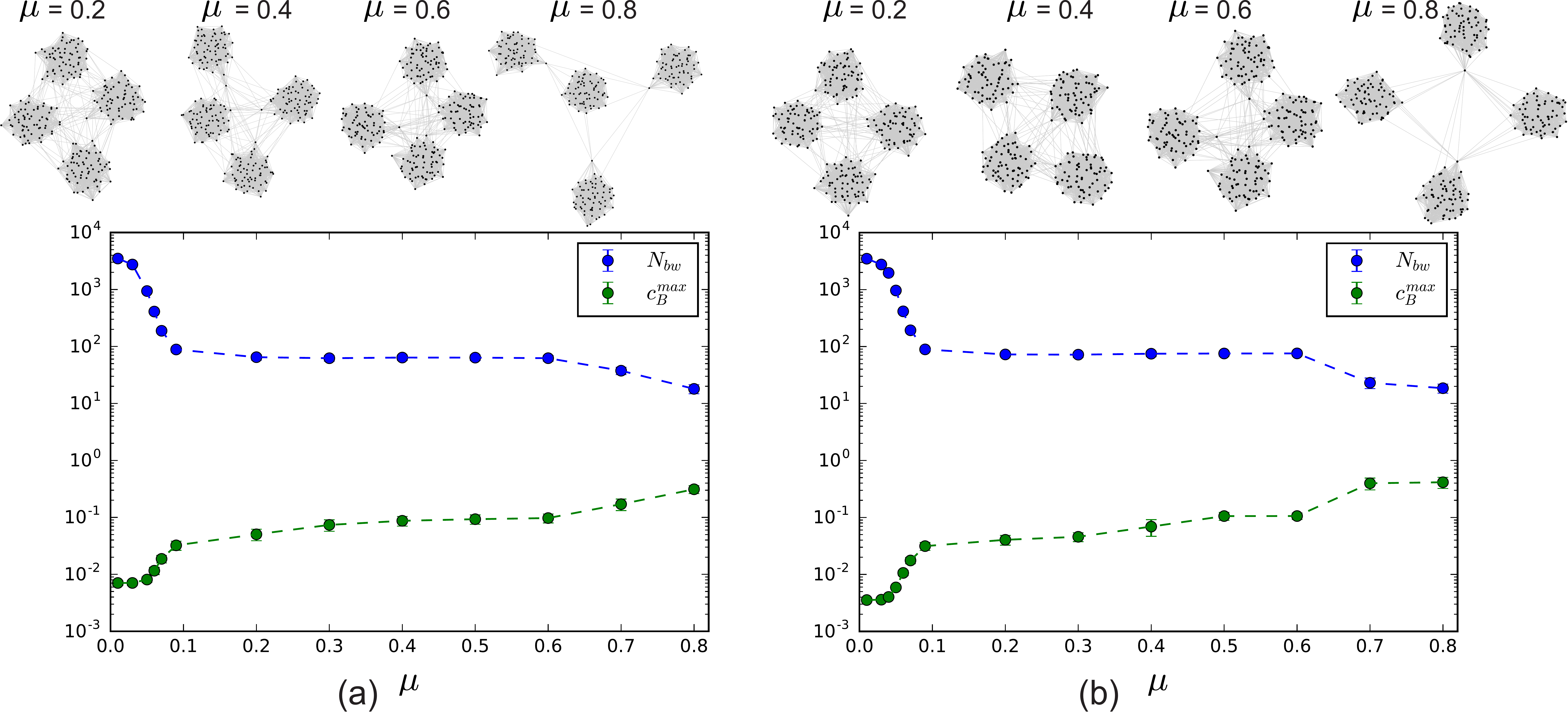}}
\caption{Rearrangement of the cluster connection pattern at the plateau exit. Dependence of the number of links between unicolor clusters versus the evolution of the maximal betweenness centrality of a node. }
\label{fig:05}
\end{figure}

The phenomena of the exit from the plateau at some critical fugacity, $\mu^{-}_{cr}$, deserves discussion. We have investigated this question via visualization of the adjacency matrix evolution. The key phenomena responsible for the plateau exit, is as follows. Consider the plateau for the two-color network. At the plateau the number of polychromatic links is stable, they connect vertices in clusters of distinct colors, and we could say that the "string" between clusters measured in number of polychromatic links, has the finite width. When we approach the value $\mu^{-}_{cr}$, the spectral density and geometry of the connections changes drastically -- see \fig{fig:05}. These changes can be easily detected by behavior of "betweenness centrality" \cite{centrality} of network nodes. The betweenness centrality of a node $v$  is given by the expression
\be
c_B=\sum_{s\ne v \ne t} \frac{\sigma_{sp}({v})}{\sigma_{sp}}
\ee
where $\sigma_{sp}$ is the total number of shortest paths from node $s$  to node $t$  and $ \sigma_{sp}(v)$ is the number of those paths that pass through $v$. Above $\mu^{-}_{cr}$ the special "gates", or "hubs" outside the clusters get formed, these hubs have very high betweenness centrality. Almost all inter-cluster connections pass through these gates. The formation of such hubs is accompanied by thinning of the string connecting different clusters.

\subsection{Semianalytic discussion of the plateau formation}

It is possible to provide the qualitative explanation of the plateau phenomena formation. To this aim consider the block Laplacian matrix $L$ for the bichromatic case
\be
L=\left(\begin{array}{cc}
A & C \\
C^{\top} & D
\end{array}\right),
\ee
and use the standard expression for the determinant of the block matrix:
\be
\det\left(\begin{array}{cc}
A & B \\
C & D
\end{array}\right) = \det(A)\; \det\left(D - CA^{-1}B\right)
\ee
We are interested in the $\mu$-dependence of first non-vanishing eigenvalue, $\lambda_2$, of Laplacian matrix, $L$:
\be
\det\left(\begin{array}{cc}
A - \lambda & C \\
C^{\top} & D -\lambda
\end{array}\right) = \det(A -\lambda)\; \det\left(D- C^{\top}(A -\lambda)^{-1}C\right) =0
\ee
The explicit form of matrices $A,D,C$ and their $\mu$-dependence is unknown however some claims concerning the plateau formation can still be done. Clearly, there are two towers of eigenvalues coming from the first and the second determinants. There is the competition between the second eigenvalue $\lambda_2^{(A)}$ of a block $A$ (note that $\lambda_1^{(A)}=0$) and the lowest eigenvalue of the second determinant. There are two possible options: (i) $\lambda_2^{(A)}$ is the lowest non-vanishing eigenvalue of the whole network, (ii) the eigenvalue  $\lambda_3^{(L)}$ of the second determinant plays the role of the algebraic connectivity. In our case the matrix of the color block $A(\mu)$ has nontrivial $\mu$-dependence, while the off-diagonal block $C$ has no explicit $\mu$-dependence in matrix elements since we have no fugacity for polychromatic links. However, the number of non-vanishing elements in $C$ is $\mu$-dependent.

\begin{figure}[ht]
\centerline{\includegraphics[width=15cm]{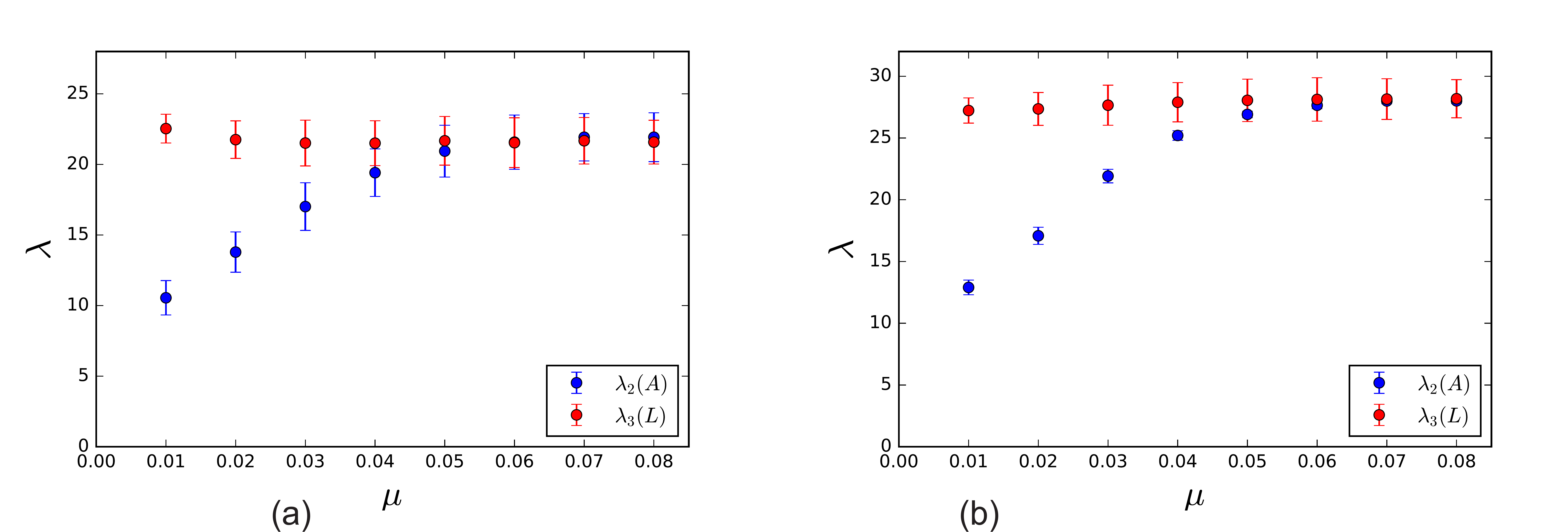}}
\caption{Dependencies of $\lambda_2^{(A)}(\mu)$ and $\lambda_3^{(L)}(\mu)$ for small values of $\mu$.}
\label{fig:07}
\end{figure}

We claim that the entrance and the exit of the plateau correspond to points of the intersection of the cluster eigenvalue $\lambda_2^A(\mu)$ and the network eigenvalue $\lambda_3^{(L)}$. The corresponding eigenvalue trajectories are presented in \fig{fig:07} where it becomes clear that $\lambda_3^{(L)}$ is almost $\mu$- independent. If $\lambda_3^{(L)}$ is the lowest non-vanishing eigenvalue of the network Laplacian matrix, there is no $\mu$-dependence in the corresponding eigenvalue. We identify this regime with plateau. The positions for the plateau entrance and exit are defined by the equation
\be
\lambda_2^{(A)}(\mu) = \lambda_3^{(L)}
\ee
Note that $\lambda_2$ at the plateau coincides with $\lambda_3^{(L)}$, whose value is close to the boundary of the continuum part of the spectrum.

The discussion of the termination of the plateau is more subtle. At the end of the plateau, the intercluster interactions proceed through hubs, which means that the matrix $C$ has the non-diagonal structure. After the exit from the plateau, one sees again the nontrivial $\mu$-dependence, which signals that the eigenvalue $\lambda_2^{(A)}(\mu)$ dominates again. The key question concerns the coherence of the layers at the plateau. The condition for the entrance at the plateau signals that the continuum parts of spectra of each cluster and of the whole network have identical supports. This means that at the plateau we have complete synchronization of layers similar to the one of \cite{arenas2}, which can be seen from the behavior of corresponding eigenvectors.

Similarly to \cite{arenas2}, we can perform the search for the algebraic connectivity by looking at the ground state of the following Hamiltonian
\be
H=\la v^{\top}|L| v \ra + g v^{\top}v +f\la v|1 \ra
\ee
where $v$ is the eigenfunction of the Laplacian while g and f are Lagrangian multipliers for the conditions of normalization of eigenvectors and the orthogonality to the homogeneous $\lambda=0$ state.  The Hamiltonian involves the Laplacian matrix of the random CERN or RRG networks hence the averaging over graphs is implied. Note that this Hamiltonian in the continuum has a lot is common with the $O(N)$-sigma model interacting with 2D gravity. Indeed the first term yields the kinetic contribution in the random geometry while the rest terms are familiar constraints in the sigma model. The phase transition in the continuum means the change of the space dependence of the ground state wave function.

\begin{figure}[ht]
\centerline{\includegraphics[width=15cm]{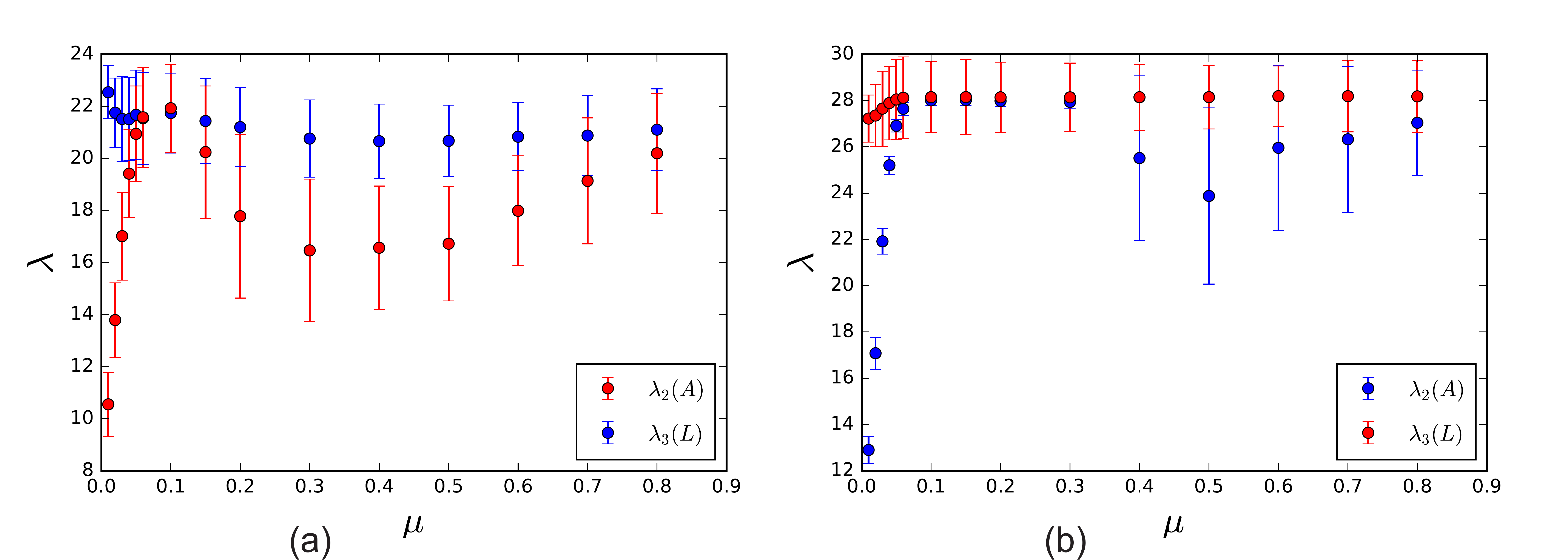}}
\caption{Dependencies of $\lambda_2^{(A)}(\mu)$ and $\lambda_3^{(L)}(\mu)$ in wide range of $\mu$.}
\label{fig:07_2}
\end{figure}

\section{Comments on localization}

It was found in \cite{anderson} that the transport properties of the perturbed CERN and RRG are nontrivial. The analysis of the level spacing distribution, $P(s)$, clearly shows that the states in the central continuum zone are delocalized, while the level spacing in the second non-perturbative zone obeys the Poissonian law, meaning that the these states are localized. Define
\be
\left\{\begin{array}{ll}
P_{deloc}(s)= A\, s\,e^{-Bs^2} & \mbox{below mobility edge, $\lambda_m$, for GOE}
\medskip \\ P_{loc}(s) = e^{-s} & \mbox{above mobility edge, $\lambda_m$, for GOE}
\end{array} \right.
\label{eq:05}
\ee
where GOE denotes the Gaussian Orthogonal Ensemble. Since delocalized modes in the main zone correspond to spectrum inside the clusters we rather figuratively associate such clusters with the metallic phase, while our entire system is a kind of an insulator. Moreover, it was found in \cite{anderson} that the spectral density in the main zone strongly depends on the graph preparation conditions, hence the corresponding delocalized states can be thought as non-ergodic.

It is worth comparing these findings with other related studies. The transport properties of the unperturbed RRG have been analyzed in \cite{spacing} where the delocalization of all modes was reported. The transport properties of the perturbed RRG with the on-site Anderson-like disorder become much more interesting. At some critical strength of the  disorder, there is the Anderson localization transition \cite{mirlin,ioffe1,ioffe2} for which the ergodic/non-ergodic behavior of delocalized states is still under discussion. The renewed interest to the one-particle localization in RRGs deals with the attempt to treat it as a model of the Fock space for some interacting many-body system \cite{levitov}. The one-particle localization in RRGs presumably is related to the phenomena of many-body localization in the real space where initial degrees of freedom are identified with the localized states in the one-particle model on the graph \cite{altshuler, gornyi}.

\begin{figure}[ht]
\centerline{\includegraphics[width=12cm]{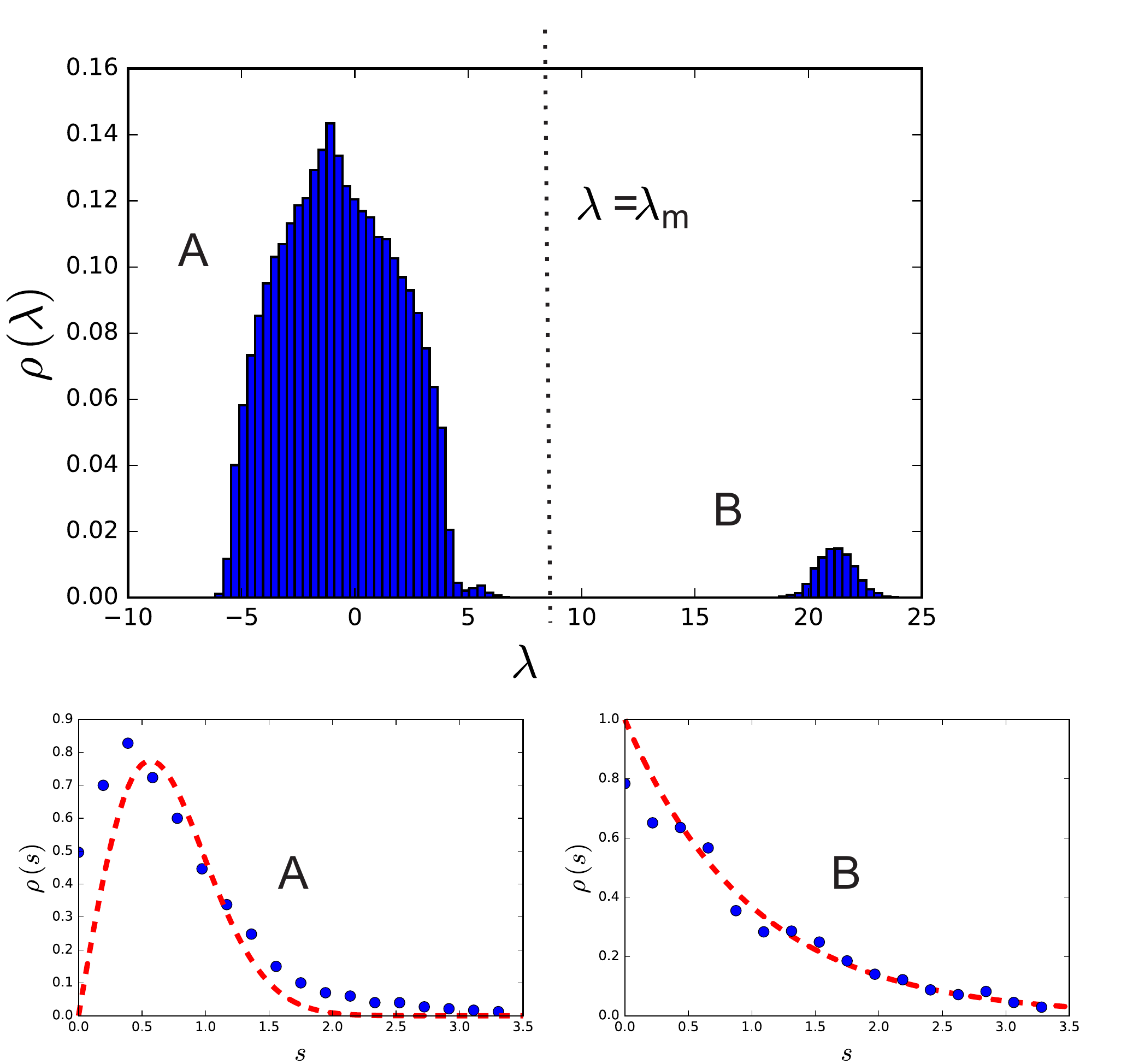}}
\caption{Level spacing distribution for 8-color RRG. Numerical results are obtained for ensemble of 250 randomly generated graphs of 256 vertices with the degree  $k=20$. }
\label{fig:08}
\end{figure}

In this Section we provide results of the similar study for polychromatic RRGs. We are focused at the level spacing distribution as a criterion of the localization. It is assumed that the number of colors is large enough to make the second zone sufficiently wide. The results are presented in the \fig{fig:08}.

It is clearly seen from \eq{eq:05} and \fig{fig:08} that the states in the main zone are delocalized while the ones if the second non-perturbative zone are localized. We can speculate that the polychromatic RRG is served as the Fock space for some interacting many-body system. Then, the one-particle localization in the polychromatic RRG is mapped onto the many-body localization in the real space and the degrees of freedom in the real space are the localized states in the Fock space.  If we accept such a point of view, we can conclude that degrees of freedom in the real space -- are clusters of different colors.

Let us highlight the new features specific for polychromatic networks and which are absent in monochromatic ones.
\begin{itemize}
\item[i)] Since the Fock space is polychromatic, one should identify the color with the specific quantum number in the initial many-body system (for instance, the color could mark the representation of some global group).
\item[ii)] In colorless networks the effects of graph topology are crucial (for example, the number of closed triangles). To the contrary, in polychromatic networks the unicolor triples of connected nodes play the key role.
\end{itemize}

We have to attribute the physical meaning to the fugacity of monochromatic triples in terms of the real space system. Since it does not involve the cycles in the Fock space, probably it could be attributed to the perturbation which mixes three energy levels. Certainly these issues deserve for further clarification.

\section{Plateau formation and exit: comparison with other models}

In \cite{color} we have conjectured that the plateau reflects the phase coexistence in the perturbed network within some interval of the fugacities, $[\mu_{cr}, \mu^{-}_{cr}]$. Below we provide qualitative remarks concerning this issue and compare it with other models. Related issue has been considered in \cite{arenas2, arenas, radicchi} in a different context. The authors considered the two-layer network constructed "by hands" from the very beginning. The parameter $p$, controlling the interlayer interaction, has been introduced and the dependence of the $\lambda_2$ on this parameter has been analyzed. At $p=0$ two layers are completely disconnected while at non-vanishing $p$ there are two regimes for $\lambda_2(p)$. At small $p$ it was found that $\lambda_2(p)\propto p$, while at $p>p^*$ the value $\lambda_2$ was independent on $p$. The plateau was assumed to be semi-infinite since no other critical behavior in $p$ is observed. It was also found in \cite{hub} that the number of hubs used to connect two layers, decreases sharply at plateau regime.

The linear connection between layers is simple enough to admit an exact solution based on the properties of block diagonal matrices with uniformly weighted off-diagonal terms. It was argued in \cite{arenas2} that the spectrum of the Laplacian matrix yielding $\lambda_2(p)$, has two regimes matching two solutions of the simple equation for the wave function overlap, if one considers the Laplacian as the Hamiltonian of the system. Denoting as $\la v_a|v_b\ra$ and $\la v_a|1 \ra$ the overlaps of parts of wave function $v=(v_a, v_b)$ (corresponding to  $\lambda_2$) with the other part and with the ground state wave function respectively. It was shown that these variables undergo the sharp transition at some $p=p^*$.  In  the analysis in \cite{arenas2,vanmighem,radicchi} the off-diagonal block of Laplacian matrix is chosen in the simplified form as  $C=p I$ while the block $A$ is $p$-independent.  To get the eigenvalues consider the determinant identity for block matrix:
\be
\det(A-\lambda +pI)\; \det \left(A-\lambda +pI - pI\frac{1}{A-\lambda +pI}pI\right)=0
\ee
Looking at the second determinant and selecting the eigenvector of $A$ with zero eigenvalue we immediately obtain the equation $(\lambda-p)^2=p^2$, which implies the competition is between the second eigenvalue, $\lambda_2^{(A)}$, of the matrix $A$ and $2p$. The plateau starts when
\be
\lambda_2^{(A)}= 2p
\ee
It was shown in \cite{arenas2} that exactly at this point the third eigenvalue, $\lambda_3^{(L)}$, hits the second one. Therefore at small values of $p$ one has $\lambda_2^{(L)}=p$ while at plateau, $\lambda_2^{(L)}=\lambda_2^{(A)}$.

Since before plateau we are dealing with the zero eigenvalue of the matrix $A$ the corresponding eigenvector is homogeneous. This allows to identify the eigenvector of the Laplacian corresponding to $2p$  which looks as
\be
V=\left(\begin{array}{cc} 1 \\ -1
\end{array}\right)
\ee
This eigenvector implies that before plateau the system can be considered as interacting  bilayer system. On the other hand at the plateau the layers are coherent and practically indistinguishable. More general phase structure with the different synchronization patterns between two layers has been analyzed in \cite{radicchi}. It depends on the ratio of the mean values of the in- and inter-layer links.

The model of \cite{arenas2} has some similarities and differences with our situation. First of all, in our case the two-layer system emerges dynamically by the spontaneous symmetry breaking and is not prepared from the very beginning. Second, we have introduced the fugacity not for inter-layer links but for the monochromatic (in-layer) trimers. However, our weight $e^{-\mu}$ plays the same role as $p$ of \cite{arenas2} since both quantities measure the relative strengths of the in-layer and inter-layer links. To compare the dependencies $\lambda_2(p)$ of \cite{arenas2} and $\lambda_2(\mu)$  of our work, is is necessary to note that small-$p$ linear regime in \cite{arenas2} corresponds to the Arrenius regime after the exit from the plateau in our case, and plateau entrance in \cite{arenas2} corresponds to the plateau exit in our case. The most striking difference between $\lambda_2(\mu)$ and $\lambda_2(p)$ is the finiteness of the plateau in the first case and seemingly infinite plateau in the last one \cite{arenas2}.  Another remark concerns the role of hubs for the inter-layer links. The exit from the plateau in our case indicates the sharp increase of the number of hubs which looks similar to the decrease of the number of hubs at the plateau entrance discussed in \cite{hub}.

To summarize, we see that the effect of discontinuity of $\lambda_2$ means the collectivization of the network layers due to the asymmetry of the relative weights of in-layer and inter-layer bonds, or more complicated motifs, like triples of monochromatic links. However, the microscopic physics behind this phenomena is still puzzling. The attempt to construct the mean-field theory of the bichromatic model is described in the Appendix.

We would like to pay attention to some similarity of the plateau emergence with the so-called Griffiths phase known in the theory of the phase transitions in disordered systems (see \cite{vojta} for review). The Griffiths phase emerges when, due to fluctuations, the spontaneous droplets are formed, in which the phase transition occurs earlier than in the whole system. To complete the phase transition in the whole system it takes some "time" to collectivize the degrees of freedom in formed clusters due to the long-range interactions. In our polychromatic networks we have ideologically similar situation. At some critical value of $\mu$, we have a kind of phase transition in the system which yields the restoration of the $Z_2$ symmetry due to the modes collectivization at the plateau above $\mu_{cr}$. However, due to the inter-layer connections, the phase transition in the whole system is incomplete and the value of the control parameter should be increased up to the termination value, $\mu^-_{cr}$, to get the new  phase in the whole system. This happens upon the plateau exit where the clusters become completely disconnected.

It is worth asking what is the continuum counterpart of the phenomena we have found in the network. Since in the continuum the network becomes the membrane and the Laplacian of the network becomes the membrane Laplacian, the natural counterpart of our network study is the interplay between the wave functions of two membrane "bubbles" associated with clusters. Indeed, the effects of collectivization of the modes in two bubbles (finite reservoirs, or clusters) takes place depending on the length and the diameter of the tube linking these bubbles \cite{Grebenkov}.

Another analogy in the continuum is as follows. Consider the quantum spectrum of the particle in the double well potential from the phase space viewpoint. If the potential is a slightly skewed, we have two minima at the classical level and two disconnected closed phase space trajectories in the $(p,x)$ plane. Quantum tunneling provides the effective connection (via instantons) between two classically disconnected regions of the phase space. From the textbooks we know what happens when the potential is $Z_2$ symmetric. Naively, we expect doubly degenerated ground state and breaking of $Z_2$ symmetry however due to the non-perturbative effects, the sum over instanton-antiinstanton pairs amounts to the level splitting and the difference between the two lowest energy levels is non-analytic in the coupling constant. From the phase space viewpoint we could consider the two regions in the phase space connected by the "non-perturbative tube". It would be interesting to pursue this analogue further.

\section{Discussion}

Our studies dealing with the generic critical behavior of perturbed CERN and RRG models have straightforward parallels in the 2D models of quantum gravity and other models involving the matrix models description with the cubic potentials and eigenvalue tunneling
\cite{marino}. In particular, in 2D gravity the eigenvalue tunneling is related to the creation of the baby-universe \cite{ambjorn} while in context of the Dijkgraaf-Vafa matrix model description of the SUSY YM, the eigenvalue tunneling corresponds to domain walls \cite{dv}. In our case the corresponding extended object is identified with the cluster, however contrary to the SYM and 2D gravity,  the immediate application of the conventional matrix model technique is impossible due to the specific properties of the ensembles of  random adjacency and Laplacian matrices and imposed constraints.

Specifically, in polychromatic networks we are dealing with the spin-like degrees of freedom on the CERN or RRG and the $M$-color case corresponds to the $M$-Potts like model. The phase transitions in the 2D gravity coupled to matter, are known as well \cite{kazakov}. The eigenvalue tunneling corresponds to the formation of the domains of the non-directed spins interacting with the random network. The action of each tunneling eigenvalue is $V(\lambda_i) -V(\lambda_0)$ where $\lambda_i$ is $i$th extremum of potential and $\lambda_0$ is the minimum of the central zone growing around. Note that in the context of the matrix model for $N=1$ SUSY YM theories this tunneling action gets identified with the tension of domain wall between the degenerated ground states. In our case "strings" between clusters seem to play the role of such domain walls hence we suggest that the effective tension of these strings is given by the instanton action.

We believe that the described critical behavior is quite universal. The whole phenomena can be formulated in physical terms as the "non-perturbative creation of insulator zone of soft modes" which could happen in any polychromatic network with the vertex degree conservation and some color- and/or topology-dependent local interaction.

At the critical value of the control parameter the network gets clusterized via the eigenvalue tunneling mechanism into the collection of weakly connected extended objects (clusters, bubbles, vesicles etc). The clusterization is accompanied by the formation of a zone of soft modes around the zero's eigenvalue of the Laplacian matrix corresponding to the whole system. These soft modes in the non-perturbative zone can be considered as the effect of collectivization of inter-cluster interactions. The hard modes live inside the clusters only and are collectivized as well.

The networks are absolutely unstable with respect to the color separation. We can consider the plateau formation for the generic multi-layer networks. Our study together with \cite{arenas2} clearly demonstrates the existence of a generic phenomena of the "layer confinement" accompanied by the collectivization of their degrees of freedom and formation of a "confined" ground state of the system. The mechanism of the ground state confinement formation is entropic and is based on an asymmetry of the inter-layer and in-layer interactions. In \cite{arenas2} the inter-layer links play the key role, while in our study the interaction of in-layer trimers results in the plateau formation. Moreover, introducing the fugacity $\mu$ to one color only (say, only to "black" trimers in bichromatic network), we shall still register the formation of two layers, and their synchronization at the plateau. This clearly demonstrates that interactions in one layer only induce the effective interactions in the second one. It is yet unclear which type of network motifs could induce the confinement of layers and which could not. In particular we see that presence of fugacities of in-layer dimers and closed triangles, is not sufficient to the confinement of layers.

We can provide simple estimates for the value of $\mu_{cr}$, at which the plateau begins. Consider one black-white bond. This bond does not contribute to energy, determined through the number of monochromatic triplets, but has a rather large entropy contribution, which can be estimated as $\sim \ln N$, where $N$ is the number of graph vertices. If we now switch this link to a single-color (for example, black) cluster, then the energy contribution associated with such a switch becomes $2\mu_{cr} Np$, because one link $(ij)$ increases the number of triples by $d_i + d_j-2$. At the plateau, the entropy contribution to free energy from this bond becomes comparable with the energy contribution in the cluster, hence we obtain the estimate for $\mu_{cr}$ for the plateau beginning, as
\be
\mu_{cr}\sim \frac{\ln N}{2 Np}
\label{eq:mu}
\ee
The expression $\eq{eq:mu}$ confirms the observation that $\mu_{cr}$ depends on $p$ as $p^{-1}$, and indicates that in the thermodynamic limit $N\to\infty$ the plateau begins at $\mu_{cr}=0$, which is consistent with the mean-field arguments presented in the Appendix A.

Recently another model involving triangles and links has been discussed in \cite{radin}. Instead of introducing the fugacity for the number of triangles, the authors fixed the numbers of triangles and links. The authors of \cite{radin} demonstrate the existence of several infinite series of the phase transitions with the different symmetry breaking patterns. It would be interesting to identify these transitions with the phase transition corresponding to a particular symmetry breaking found in \cite{decay}. Note that in our study there are quite complicated symmetry breaking states at the intermediate stages of the stochastic evolution. However the symmetry breaking in the ground state in perturbed CERN and RRG is unique. This suggests that possibly many phases found in microcanonical ensemble in \cite{radin} correspond to the intermediate metastable states in our mixed ensemble.

The striking similarity of the color separation in polychromatic networks and the QCD chiral symmetry breaking seems very challenging. In the QCD case the instanton-antiinstanton ensemble is considered and the vertex degree which counts the number of instanton's zero modes remains unchanged, being topologically fixed. The overlap matrix for zero modes plays the role of the Laplacian matrix for the network. Playing with some control parameter, we can get the formation about extended objects built from collection of instantons. Since graph Dirac operator can be considered, roughly speaking, as the square root of the graph Laplacian matrix, the behavior of the spectrum of the Laplacian matrix near vanishing eigenvalue can tell us about the corresponding behavior of the spectrum of the Dirac operator. The latter is relevant for the chiral condensate formation. The existence of the non-perturbative soft zone around $\lambda=0$, which we have found for the Laplacian, yields the condensate emergence via the Casher-Banks relation for the Dirac operator spectrum $\la\bar{\Psi}\Psi\ra = - \pi \rho(0)$. We have found in this paper that states in the soft part of the spectrum are localized and appearance of the non-perturbative zone  corresponds to the Anderson transition. This rhymes with the interpretation of the chiral phase transition in QCD as the Anderson transition \cite{koscacz}. It would be very interesting to develop this line of reasoning further.

The spectral gap can been considered as the order parameter in the statistical systems. Recently it was argued that the collision of two largest eigenvalues of the entanglement Hamiltonian for two parts of the whole system in the Schmidt decomposition, corresponds exactly to the point of the phase transition in the whole system.  Moreover the spectrum of the entanglement Hamiltonian
defines the entanglement entropy. In the context of the tensor networks the entanglement entropy of two subnetworks is closely
related to the number of links to be cut to make two subnetworks disconnected. This certainly has parallels with our study
and we hope to discuss the spectrum of the entanglement Hamiltonian in bichromatic and polychromatic models and entanglement
entropy of the color clusters in a separate publication.

We are grateful for D. Grebenkov and M. Tamm for the useful discussions. The work of A.G. was performed at the Institute for Information Transmission Problems with the financial support of the Russian Science Foundation (Grant No.14-50-00150); S.N. acknowledges the support of the EU-Horizon 2020 IRSES project DIONICOS (612707), and of the RFBR grant No. 16-02-00252.

\begin{appendix}

\section{Partition function of bichromatic network in the mean-field approximation}

Let us provide the mean-field-like consideration of the bichromatic network with fixed fugacity of unicolor trimers (irrespective of their topology).  Introduce $s_{ij}=0,1,-1$ the "spin" variable corresponding to the bond $ij$ (the matrix element $a_{ij}$ in the adjacency matrix $A$):
\be
s_{ij}=\left\{\begin{array}{cl} +1 & \mbox{bond between same colors} \\ -1 & \mbox{bond between different colors} \\ 0 & \mbox{absence of a bond} \end{array} \right.
\ee
The Hamiltonian $H_{ijk}$ of the interactions between monochromatic triples $(ijk)$ of network bonds reads:
\be
H_{ijk}=\frac{\mu}{4}\left(s_{ij} s_{jk} + s_{ij} s^2_{jk} + s^2_{ij} s_{jk} + s^2_{ij} s^2_{jk}
\right)
\ee
The Hamiltonian $H_{ijk}$ is equal to $\mu$ if and only if the neighboring values of $s_{ij}, s_{jk}$ correspond to the same color. If they correspond to different colors, or are absent, then $H_{ijk}=0$.

The partition function depending on the concentration of black-white bonds, $c_{bw}$, can be written as follows
\be
Z(c_{bw}) = \sum_{\{s_{ij}=0,\pm 1\}}\exp\left(\sum_{ijk}^N H_{ijk}\right)
\delta\left(\sum_{ijk}^N\left(s_{ij}-s^2_{ij}+2c_{bw}\right)\right)
\prod_{j=1}^N\delta\left(\sum_{i=1}^Ns^2_{ij}-d_j\right)
\label{pf}
\ee
where $d_j$ is the vertex degree of the graph (if $d_j=d$ for all $j=1,...,N$, then we have a regular random graph). The partition function \eq{pf} is the exact expression for requested partition function of two-color network.

Let us transform the terms in the Hamiltonian $H_{ijk}$ in the following way
\be
\begin{array}{rcl}
\frac{\mu}{4} s_{ij}s_{jk} & = & \frac{\mu}{8}\left(s_{ij} + s_{jk}\right)^2 - \frac{\mu}{8}\left(s^2_{ij} + s^2_{jk}\right) \medskip \\
\frac{\mu}{4} s_{ij}s^2_{jk} & = & \frac{\mu}{8}\left(s_{ij} + s^2_{jk}\right)^2 - \frac{\mu}{8}\left(s^2_{ij} + s^4_{jk}\right) \medskip \\
\frac{\mu}{4} s^2_{ij}s_{jk} & = & \frac{\mu}{8}\left(s^2_{ij} + s_{jk}\right)^2 - \frac{\mu}{8} \left(s^4_{ij} + s^2_{jk}\right) \medskip \\ \frac{\mu}{4} s^2_{ij}s^2_{jk} & = & \frac{\mu}{8}\left(s^2_{ij} + s^2_{jk}\right)^2 - \frac{\mu}{8} \left(s^4_{ij} + s^4_{jk}\right)
\end{array}
\label{sq}
\ee
Now we can introduce the auxiliary Gaussian fields which allow to decouple interacting terms in
\eq{sq}:
\be
\begin{array}{rcl}
\disp e^{\frac{\mu}{8}\left(s_{ij} + s_{jk}\right)^2} & = & \disp \sqrt{\frac{2}{\mu\pi}} \int_{-\infty}^{\infty} d\varphi_{ijk}\; e^{-\frac{2}{\mu}\varphi_{ijk}^2 + \left(s_{ij} + s_{jk}\right)\varphi_{ijk}} \medskip \\
\disp e^{\frac{\mu}{8}\left(s_{ij} + s^2_{jk}\right)^2} & = & \disp \sqrt{\frac{2}{\mu\pi}} \int_{-\infty}^{\infty} d\chi_{ijk}\; e^{-\frac{2}{\mu}\chi_{ijk}^2 + \left(s_{ij} + s^2_{jk}\right)\chi_{ijk}} \medskip \\
\disp e^{\frac{\mu}{8}\left(s^2_{ij} + s_{jk}\right)^2} & = & \disp \sqrt{\frac{2}{\mu\pi}}	\int_{-\infty}^{\infty} d\omega_{ijk}\; e^{-\frac{2}{\mu}\omega_{ijk}^2 + \left(s^2_{ij} + s_{jk}\right)\omega_{ijk}} \medskip \\
\disp e^{\frac{\mu}{8}\left(s^2_{ij} + s^2_{jk}\right)^2} & = & \disp \sqrt{\frac{2}{\mu\pi}}	\int_{-\infty}^{\infty} d\psi_{ijk}\; e^{-\frac{2}{\mu}\psi_{ijk}^2 + \left(s^2_{ij} + s^2_{jk}\right)\psi_{ijk}}
\end{array}
\label{hub}
\ee
To each triad of vertices $ijk$ we identify for independent Gaussian fields $\varphi_{ijk}, \chi_{ijk},\omega_{ijk}, \psi_{ijk}$. Having such a representation (in fact, this is a sort of a Hubbard-Stratonovich transform), we decouple adjacent spins $s_{ij}$ and $s_{jk}$ interacting in the common graph vertex $j$. Exponentiating all the Kronecker $\delta$-functions in \eq{pf},
\be
\delta(x) = \frac{1}{2\pi i} \oint \frac{d\lambda}{\lambda^{x+1}} = \begin{cases} 1 & \mbox{for	$x=0$} \\ 0 & \mbox{otherwise} \end{cases}
\ee
and using \eq{hub}, may rewrite the partition function \eq{pf} as follows:
\begin{multline}
Z(c_{bw},d_1,...,d_N) = \left(\frac{2}{\mu\pi}\right)^{6N} \frac{1}{(2\pi i)^{N+1}}
\oint d\lambda\; \lambda^{-2c_{bw}-1} \prod_{j=1}^N \oint d\xi_j\; \xi_j^{d_j-1} \\
\times \prod_{ijk}^N \int\limits_{-\infty}^{\infty} d\varphi_{ijk}\;
e^{-\frac{2}{\mu}\varphi_{ijk}^2} \int\limits_{-\infty}^{\infty} d\chi_{ijk}\;
e^{-\frac{2}{\mu}\chi_{ijk}^2} \int\limits_{-\infty}^{\infty} d\omega_{ijk}\;
e^{-\frac{2}{\mu}\omega_{ijk}^2} \int\limits_{-\infty}^{\infty} d\psi_{ijk}\;
e^{-\frac{2}{\mu}\psi_{ijk}^2}
\\ \times \sum_{\{s_{ij}=0,\pm 1\}}\exp\Bigg\{\sum_{ijk}^N \Big[\left(s_{ij} +
s_{jk}\right)\varphi_{ijk} + \left(s_{ij} + s^2_{jk}\right)\chi_{ijk} + \left(s^2_{ij} +
s_{jk}\right)\omega_{ijk} + \left(s^2_{ij} + s^2_{jk}\right)\psi_{ijk} \\ -
\frac{\mu}{4}\left(s^2_{ij} + s^2_{jk}\right)-\frac{\mu}{4}\left(s^4_{ij} +
s^4_{jk}\right)-\left(s_{ij}-s^2_{ij} \right) \ln \lambda - s_{ij}^2 \ln \xi_j\Big]\Bigg\}
\label{pf2}
\end{multline}
In \eq{pf2} all spins $s_{ij}$ sitting on bonds are decoupled and hence we can first perform summation over spins $\{s_{ij}=0,\pm 1\}$ to get an effective action for Gaussian fields $\varphi_{ijk}$ and $\psi_{ijk}$. We have used the following identities (since the summation is carried out over the whole network and all bonds are independent):
\be
\begin{array}{l}
\disp \prod_{ijk} \exp\left[(s_{ij}+s_{jk})\varphi_{ijk}\right] \equiv \prod_{ijk} \exp\left[2s_{ij}\,\varphi_{ijk}\right] = \prod_{ij}\exp\left[2s_{ij}\sum_{k=1}^N \varphi_{ijk}\right] \medskip \\
\disp \prod_{ijk} \exp\left[(s_{ij}+s^2_{jk})\chi_{ijk}\right] \equiv \prod_{ijk} \exp\left[(s_{ij}+s^2_{ij})\chi_{ijk}\right] = \prod_{ij}\exp\left[(s_{ij}+s^2_{ij})\sum_{k=1}^N \chi_{ijk}\right]
\end{array}
\ee
and similar identities for the fields $\omega_{ijk}$ and $\psi_{ijk}$.

For summation over spins on bonds we suppose that the each link could take 3 independent values: $s_{ij}=0,+1,-1$.  Performing summation, we arrive at the following expression for the effective partition function:
\begin{multline}
Z(c_{bw},d_1,...,d_N) = \left(\frac{2}{\mu\pi}\right)^{2N} \frac{1}{2\pi i} \oint d\lambda\; \lambda^{-2c_{bw}-1} \prod_{j=1}^N \frac{1}{2\pi i} \oint d\xi_j\; \xi_j^{d_j-1}  \\
\prod_{ijk}^N\int\limits_{-\infty}^{\infty}...\int\limits_{-\infty}^{\infty} d\varphi_{ijk} d\chi_{ijk} d\omega_{ijk} d\psi_{ijk}\; e^{-\frac{2}{\mu}\left(\varphi_{ijk}^2+ \chi^2_{ijk}+ \omega^2_{ijk}+\psi^2_{ijk}\right)} \\
\times \prod_{ijk} \Bigg\{1+e^{-\mu N-\ln \xi_j}\Bigg[\exp\left(2\sum_{k=1}^N\left(\varphi_{ijk}+\chi_{ijk}+\omega_{ijk}+
\psi_{ijk}\right)\right) \\ + \exp\left(-2\sum_{k=1}^N\left(\varphi_{ijk}-\psi_{ijk}\right)-2\ln
\lambda\right) \Bigg]\Bigg\}
\end{multline}

The effective Hamiltonian, $H$, reads
\begin{multline}
H=-\frac{2}{\mu} \sum_{ijk} \left(\varphi_{ijk}^2+ \chi^2_{ijk}+ \omega^2_{ijk}+\psi^2_{ijk} \right) \\
+ \sum_{ij}\ln\Bigg\{1+e^{-\mu N-\ln\xi_j}\Bigg[\exp\left(2\sum_{k=1}^N\left(\varphi_{ijk}+\chi_{ijk}+\omega_{ijk}+ \psi_{ijk}\right)\right) \\ + \exp\left(-2\sum_{k=1}^N\left(\varphi_{ijk}-\psi_{ijk}\right)-2\ln \lambda\right) \Bigg] \Bigg\}
\end{multline}

The minimization of $H$ over all fields leads to the following system of equations:
\be
\left\{\begin{array}{l}
\disp \frac{\partial H}{\partial \varphi_{ijk}} = 0 \medskip \\
\disp \frac{\partial H}{\partial \chi_{ijk}} = 0 \medskip \\
\disp \frac{\partial H}{\partial \omega_{ijk}} = 0 \medskip \\
\disp \frac{\partial H}{\partial \psi_{ijk}} = 0
\end{array} \right.
\quad \Rightarrow \quad \left\{\begin{array}{l} \disp \frac{4}{\mu} \varphi_{ijk} - \frac{e^{-\mu N-\ln\xi_j}\left[e^{2\sum_k(\varphi_{ijk}+\chi_{ijk}+\omega_{ijk}+ \psi_{ijk})}- e^{-2\sum_k(\varphi_{ijk}-\psi_{ijk})-2\ln \lambda}\right]}{A_{ij}} = 0 \medskip \\
\disp \frac{4}{\mu} \chi_{ijk} - \frac{e^{-\mu N-\ln\xi_j}\; e^{2\sum_k(\varphi_{ijk}+\chi_{ijk}+\omega_{ijk}+\psi_{ijk})}}{A_{ij}} = 0 \medskip \\
\disp \frac{4}{\mu} \omega_{ijk} -  \frac{e^{-\mu N-\ln\xi_j}\; e^{2\sum_k(\varphi_{ijk}+\chi_{ijk}+ \omega_{ijk}+\psi_{ijk})}}{A_{ij}} = 0 \medskip \\
\disp \frac{4}{\mu} \psi_{ijk} - \frac{e^{-\mu N-\ln\xi_j}\left[e^{2\sum_k(\varphi_{ijk}+\chi_{ijk}+\omega_{ijk}+\psi_{ijk})}+ e^{-2\sum_k(\varphi_{ijk}-\psi_{ijk})-2\ln \lambda}\right]}{A_{ij}} = 0
\end{array} \right.
\label{all}
\ee
where
\begin{multline}
A_{ij}=1+e^{-\mu N-\ln\xi_j}\Bigg[\exp\left(2\sum_{k=1}^N\left(\varphi_{ijk}+\chi_{ijk}+\omega_{ijk}+ \psi_{ijk}\right)\right) \\ + \exp\left(-2\sum_{k=1}^N\left(\varphi_{ijk}-\psi_{ijk}\right)-2\ln \lambda\right) \Bigg]
\label{a}
\end{multline}

From \eq{all} we see that $\chi_{ijk} = \omega_{ijk}$. Let us introduce now two new composite fields
\be
\begin{array}{l}
\disp u_{ij} = \sum_k(\varphi_{ijk}+\chi_{ijk}+\omega_{ijk}+\psi_{ijk}) \medskip \\
\disp v_{ij} = \sum_k(\varphi_{ijk}-\psi_{ijk})
\end{array}
\ee
In terms of $u_{ij}$ and $v_{ij}$ we can rewrite \eq{all}--\eq{a} as a system of two independent
equations:
\be
\left\{\begin{array}{l} \disp \frac{1}{\mu N} u_{ij} - \frac{e^{-\mu N-\ln\xi_j} e^{2u_{ij}}}{1+e^{-\mu N-\ln\xi_j}\left(e^{2u_{ij}}+e^{-2v_{ij}-2\ln \lambda}\right)} = 0 \medskip \\
\disp \frac{2}{\mu N} v_{ij} + \frac{e^{-\mu N-\ln\xi_j-2\ln \lambda-2v_{ij}}} {1+e^{-\mu N-\ln\xi_j}\left(e^{2u_{ij}}+e^{-2v_{ij}-2\ln \lambda}\right)}=0
\end{array} \right.
\ee

In the mean-field approximation we suggest that $u_{ij}\equiv u$, $v_{ij}\equiv v$, and $\xi_j\equiv \xi$, and obtain the closed system of two equations on two fields $u,v$:
\be
\left\{\begin{array}{l} \disp \frac{1}{\mu'} u - \frac{e^{-\mu'-\ln\xi} e^{2u}}{1+e^{-\mu'-\ln\xi}\left(e^{2u}+e^{-2v-2\ln \lambda}\right)} = 0 \medskip \\
\disp \frac{2}{\mu'} v + \frac{e^{-\mu'-\ln\xi-2\ln\lambda-2v}} {1+e^{-\mu'-\ln\xi}\left(e^{2u}+e^{-2v-2\ln \lambda}\right)}=0
\end{array} \right.
\label{a:eqs}
\ee
where we have introduced the normalized chemical potential of trimers, $\mu' = \mu N$.

\begin{figure}[ht]
\centerline{\includegraphics[width=10cm]{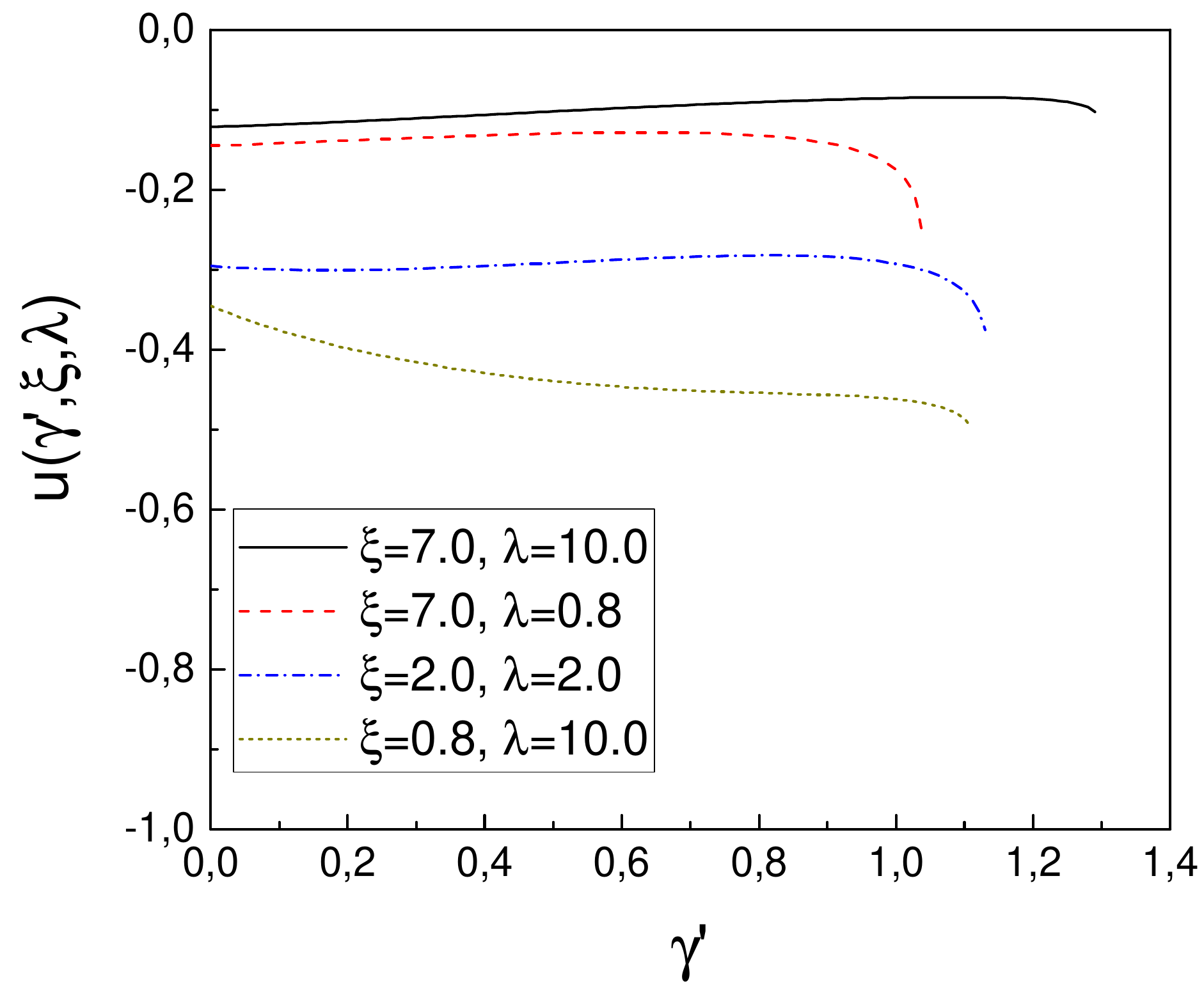}}
\caption{Depencence $u(\mu')$ for few values of fixed parameters, $\xi$ and $\lambda$. In a broad interval of these parameters, the value $u$ is almost independent on $g$.}
\label{a:plateau}
\end{figure}

It turns out that the system \eq{a:eqs} can be converted to the single transcendental equation for one unknown function, $u$ (respectively,  $v$). The equation for $u$ reads
\be
u-\lambda^2\left(2\mu'-u\left(\xi e^{\mu'-2u}+1\right)\right)\exp\left(3u+u\xi e^{\mu'-2u}-2\mu'\right)=0
\label{a:eq2}
\ee
The \eq{a:eq2} can be analyzed in the regime when $|u|\ll 1$. Expanding \eq{a:eq2} up to quadratic terms in $u$ and solving corresponding algebraic equation in a form $u=u(\xi,\lambda,\mu')$, we can identify the regime when the $u$ does no depend (or very weakly depends) on $\mu'$. This regime would correspond to the emergence of the plateau. In the figure \fig{a:plateau} we have shown the dependence $u(\mu')$ for few fixed values of $\xi$ and $\lambda$.

For rather broad interval of $\xi$ and $\lambda$, we see that $u(\mu')\approx \mathrm{const}$, which could be regarded as a strong evidence of plateau existence, since the dependence $u(\mu')$ can be straightforwardly translated to the dependence of the concentration of black-white bonds on $\mu'$. The proposed analysis is very preliminary and deserves more involved consideration, however first indications that the effect of the plateau formation can be catched within the mean-field consideration make us pretty optimistic.

\end{appendix}

\end{document}